# A large scale dynamo and magnetoturbulence in rapidly rotating core-collapse supernovae

Philipp Mösta[1,2], Christian D. Ott[1], David Radice[1], Luke F. Roberts[1], Erik Schnetter[3,4,5], & Roland Haas[6]

[1]*TAPIR, Walter Burke Institute for Theoretical Physics, Mailcode 350-17, California Institute of Technology, Pasadena, CA 91125, USA, pmoesta@tapir.caltech.edu*

[2]*Department of Astronomy, 501 Campbell Hall #3411, University of California at Berkeley, Berkeley, CA 94720*

[3]*Perimeter Institute for Theoretical Physics, Waterloo, ON, Canada*

[4]*Department of Physics, University of Guelph, Guelph, ON, Canada*

[5]*Center for Computation & Technology, Louisiana State University, Baton Rouge, LA, USA*

[6]*Max Planck Institute for Gravitational Physics, Am Mühlenberg 1, 14476 Potsdam-Golm, Germany*

**Magnetohydrodynamic (MHD) turbulence is of key importance in many high-energy astrophysical systems, where MHD instabilities can amplify local magnetic field over very short time scales[1,2]. Specifically, the magnetorotational instability (MRI) and dynamo action[3,4] have been suggested as a mechanism to grow magnetar-strength magnetic field ($\geq 10^{15}$ G) and magnetorotationally power the explosion[5-7] of a rotating massive star[8,9]. Such stars are progenitor candidates for type Ic-bl hypernova explosions[10,11] and make up all supernovae connected to long**

**gamma-ray bursts (GRBs)[12,13]. The MRI has been studied with local[14,15,16] high-resolution shearing box simulations in 3D or with global 2D simulations[17], but it is an open question whether MRI-driven turbulence can result in the creation of a large-scale ordered and dynamically relevant field. Here we report results from global 3D general-relativistic magnetohydrodynamic (GRMHD) turbulence simulations and show that MRI-driven MHD turbulence in rapidly rotating protoneutron stars produces an inverse cascade of energy. We find a large-scale ordered toroidal field that is consistent with the formation of bipolar magnetorotationally driven outflows. Our results demonstrate that rapidly rotating massive stars are plausible progenitors for both type Ic-bl supernovae[10,18,19] and long GRBs, present a viable formation scenario for magnetars[20,21], and may account for potentially magnetar-powered superluminous supernovae[22,23].**

We study MHD turbulence in the shear layer around a rapidly rotating protoneutron star using high-resolution (~10 times higher than previous simulations) global 3D GRMHD simulations. We take initial conditions from a full 3D GRMHD adaptive mesh refinement (AMR) simulation[7] of stellar collapse in a rapidly spinning progenitor star (with initial spin period of the iron core $P_0 = 2.25$ s before collapse, spin period of the protoneutron star after core bounce $P_{PNS} = 1.18$ ms, and initial maximum magnetic field $10^{10}$ G) at $t_{map} = 20$ ms after core bounce. At this time linear winding[24] has built up maximum toroidal field of $\simeq 7 \cdot 10^{14}$ G close to the rotation axis of the protoneutron star and $\simeq 3 \cdot 10^{14}$ G in the equatorial region. The maximum poloidal magnetic field is $\simeq 7 \cdot$ $10^{14}$ G at $t_{map}$ = 20 ms after core bounce. We carry out simulations in four resolutions, dx = [500 m, 200 m, 100 m, 50 m], adopt a domain size of 66.5 km in x and y direction and 133 km in z direction (rotation axis), and employ a 90° rotational symmetry in the xy-plane (no symmetry in z). This allows us to study the MRI-unstable layer surrounding the core of the protoneutron star with unprecedented resolution with fully self-consistent global 3D simulations of MHD turbulence in stellar collapse.

The two lowest resolution simulations show no or only minor toroidal magnetic field amplification consistent with not resolving the fastest growing mode (FGM) of the MRI. The toroidal field in the two highest resolution simulations exhibits exponential growth soon after the start of our simulations (Fig. 1). The poloidal magnetic field evolution follows the toroidal one closely (Extended Data Fig. 4). The initial transition to exponential growth in both the global maximum toroidal field (panel (a) Fig. 1) and the maximum toroidal field in a box with height 7.5 km above and below the equatorial plane (panel (b) Fig. 1) is nearly identical between the 100 m and 50 m simulations. This indicates that we resolve the FGM of the MRI with the 100 m simulation and is consistent with our background flow stability analysis of the initial AMR simulation (Extended Data Fig. 3). The observed growth rate of $\tau \simeq 0.5$ ms agrees well with the analytically predicted growth rate of the FGM from linear analysis. The field evolution quickly becomes non-linear and this rapid growth reaches a fully turbulent saturated state within 3 ms. The turbulent saturated toroidal field strength agrees to within a factor of two between the two highest resolution simulations (100 m and 50 m). Once non-linear field strength is reached, secondary modes and couplings between individual modes become important for the observed growth rate of the MRI. The final turbulent saturation field is not converged and differs between resolutions because secondary instabilities, resistivity, and finite resolution effects become important[25,26]. However, these differences decrease with increasing resolution and we expect our results to hold when even higher-resolution simulations become computationally accessible. This is supported by the fact that the local features of our global 3D simulations are consistent with previous higher resolution (dx ≃ 10 m) local simulations of the MRI[14].

The resolution dependence of the magnetic field in the turbulent state is striking (Fig. 2). While the 500 m and 200 m simulations show none to only mild turbulence, the 100 m and 50 m simulations develop a fully turbulent shear layer around the protoneutron star. We observe radial filaments of magnetic field that oscillate from negative to positive values on a length scale of 1 km, consistent with the predicted wavelength of the FGM of the MRI (Extended Data Fig. 3). These structures resemble channel flow formation observed in shearing box simulations[14] but do not stay coherent due to the background flow. Similar, non-coherent filaments were also observed in the 2D global simulations of [17].

The turbulent kinetic and electromagnetic energy spectra calculated from our simulations are shown in Fig. 3. Initially, the turbulent kinetic energy, which is nearly constant in time, is several orders of magnitude larger across all scales than the electromagnetic energy. The electromagnetic energy is highly time and resolution dependent. While the low resolution calculation shows little evolution away from the initial spectrum, the higher resolution calculations saturate at larger and larger energy at large k (panel (a) Fig. 3). The saturation value at large and intermediate k is within a factor of 3 of equipartition with the turbulent kinetic energy in the 50 m calculation. After saturation is reached at large k, we observe an inverse cascade of energy triggering growth of large-scale electromagnetic energy peaked at k = 4, which corresponds to a length scale of 5 km for our domain. This is well below the driving scale of the FGM of the MRI (k ≈ 20) and consistent with the structures evident in panel (d) of Fig. 2 and panel (c) of Fig. 4. The growth in the first 7 ms is fitted well by an exponential with e-folding time $\tau$ = 3.5 ms. We observe a transition away from clean exponential growth for t - $t_{map}$ ≥ 7 ms, which may be caused by the magnetic field becoming dynamically relevant and/or (numerical) resistivity becoming important for the magnetic field evolution[15]. Here, the growth at k = 4 is better described by a linear fit. In an inverse cascade the energy is expected to reach approximately the same relative saturation value (with respect to the driving turbulent kinetic energy) at all k's with sufficiently long evolution time[3,4]. We find evidence for this in the range 10 ≤ k ≤ 50 where the magnetic energy spectrum begins to evolve towards a similar power-law scaling as the turbulent kinetic energy. Assuming this holds also at smaller k, we extrapolate the growth of magnetic energy based on the linear fit (panel (c), Fig. 3). We expect to reach saturation electromagnetic energy at small k within t - $t_{map}$ ≈ 60 ms. The observed difference between the 100 m and 50 m resolution calculations in the saturation energy at large k and in the inverse energy cascade indicates that the turbulent state is not fully captured with the 100 m simulation and that the efficiency of the inverse cascade may still increase when going to even higher resolution than 50m.

Our results indicate that the electromagnetic energy will rival the turbulent kinetic energy and dominate the less efficient neutrino heating[6,27]. Therefore MHD stresses are likely the dominant factor in reviving the stalled shock in rapidly rotating progenitors. Furthermore, we observe formation of large-scale structured toroidal magnetic field near the rotation axis of the protoneutron star in the later stages of the 50 m simulation (panel (c), Fig. 4, panel (d) Extended Data Fig. 6). This large-scale field is not present in the initial data (panel (a), Fig. 4), nor does it develop in the lower resolution cases (panel (b), Fig. 4, panels (a)-(c) Extended Data Fig. 6). This magnetar-strength toroidal field close to the rotation axis is a strong indication that hoop stresses, which favour the formation of MHD-powered outflows are present along the poles[5,24]. Velocity vectors along the rotation axis are pointing outwards towards the end of the 50 m simulation (Extended Data Fig. 7). Our findings have significant implications for stellar collapse in rapidly rotating massive stars. The MRI is a weak-field instability (i.e. its growth rate $\tau_{MRI}$ does not depend on the strength of the magnetic field) and the observed rapid e-folding time of $\tau \simeq 0.5$ ms is short enough such that the scenario presented here is viable even for much weaker initial seed fields. In addition, the MRI was shown to operate efficiently in purely toroidal, mixed poloidal/toroidal and random magnetic field configurations [2]. Hence, we expect our results to hold for arbitrary precollapse magnetic field configurations. Additionally, m = 1 instabilities, shown to alter the explosion geometry of jet explosions in the full 3D simulations of [7], will start to become relevant only after large-scale toroidal field of magnetar strength has been built up (the instability criterion depends on having an ultra-strong toroidal field present in the first place[7]). This makes MHD-driven explosions a likely scenario in rapidly rotating progenitors independent of the initial magnetisation of the star with the explosion geometry likely being of the double-lobe form shown in[7]. Additionally, the large-scale build up of magnetic field in the shear layer of the protoneutron star demonstrates that MRI-driven turbulence is a promising mechanism to form pulsars and magnetars in rapidly rotating stellar collapse. This indicates that rapidly rotating massive stars can also account for potentially magnetar-powered superluminous supernovae[22,23].

**Online Content** Methods, along with any additional Extended Data display items and Source Data, are available in the online version of the paper; references unique to these sections appear only in the online paper.

**References**

1       Fricke, K. Stability of Rotating Stars II. The Influence of Toroidal and Poloidal Magnetic Fields. *Astronomy and Astrophysics* **1**, 388 (1969).
2       Balbus, S. A. & Hawley, J. F. A powerful local shear instability in weakly magnetized disks. I - Linear analysis. II - Nonlinear evolution. *The Astrophysical Journal* **376**, 214-233 (1991).
3       Frisch, U., Pouquet, A., Leorat, J. & Mazure, A. Possibility of an inverse cascade of magnetic helicity in magnetohydrodynamic turbulence. *Journal of Fluid Mechanics* **68**, 769-778 (1975).
4       Moffatt, H. K. *Magnetic field generation in electrically conducting fluids*. (1978).
5       LeBlanc, J. M. & Wilson, J. R. A Numerical Example of the Collapse of a Rotating Magnetized Star. *The Astrophysical Journal* **161**, 541 (1970).
6       Burrows, A., Dessart, L., Livne, E., Ott, C. D. & Murphy, J. Simulations of Magnetically Driven Supernova and Hypernova Explosions in the Context of Rapid Rotation. *The Astrophysical Journal* **664**, 416-434 (2007).
7       Mösta, P. *et al.* Magnetorotational Core-collapse Supernovae in Three Dimensions. *The Astrophysical Journal Letters* **785**, L29 (2014).
8       Akiyama, S., Wheeler, J. C., Meier, D. L. & Lichtenstadt, I. The Magnetorotational Instability in Core-Collapse Supernova Explosions. *The Astrophysical Journal* **584**, 954-970 (2003).

9       Thompson, T. A., Quataert, E. & Burrows, A. Viscosity and Rotation in Core-Collapse Supernovae. *The Astrophysical Journal* **620**, 861-877 (2005).

10      Drout, M. R. *et al.* The First Systematic Study of Type Ibc Supernova Multi-band Light Curves. *The Astrophysical Journal* **741**, 97 (2011).

11      Soderberg, A. M. *et al.* Relativistic ejecta from X-ray flash XRF 060218 and the rate of cosmic explosions. *Nature* **442**, 1014-1017 (2006).

12      Hjorth, J. & Bloom, J. S. in *Chapter 9 in "Gamma-Ray Bursts", Cambridge Astrophysics Series 51, eds. C. Kouveliotou, R. A. M. J. Wijers and S. Woosley, Cambridge University Press (Cambridge), p. 169-190*     169-190 (2012).

13      Modjaz, M. Stellar forensics with the supernova-GRB connection. *Astronomische Nachrichten* **332**, 434-447 (2011).

14      Obergaulinger, M., Aloy, M. A. & Müller, E. Axisymmetric simulations of magneto-rotational core collapse: dynamics and gravitational wave signal. *Astronomy and Astrophysics* **450**, 1107-1134 (2006).

15      Brandenburg, A. & Subramanian, K. Astrophysical magnetic fields and nonlinear dynamo theory. *Physics Reports* **417**, 1-209 (2005).

16      Masada, Y., Takiwaki, T. & Kotake, K. Magnetohydrodynamic Turbulence Powered by Magnetorotational Instability in Nascent Protoneutron Stars. *The Astrophysical Journal Letters* **798**, L22 (2015).

17      Sawai, H., Yamada, S. & Suzuki, H. Global Simulations of Magnetorotational Instability in the Collapsed Core of a Massive Star. *The Astrophysical Journal Letters* **770**, L19 (2013).

18      Galama, T. J. *et al.* An unusual supernova in the error box of the $\gamma$-ray burst of 25 April 1998. *Nature* **395**, 670-672 (1998).

19      Woosley, S. E. & Heger, A. The Progenitor Stars of Gamma-Ray Bursts. *The Astrophysical Journal* **637**, 914-921 (2006).

20      Thompson, C. & Duncan, R. C. Neutron star dynamos and the origins of pulsar magnetism. *The Astrophysical Journal* **408**, 194-217 (1993).

21      Duncan, R. C. & Thompson, C. Formation of very strongly magnetized neutron stars - Implications for gamma-ray bursts. *The Astrophysical Journal Letters* **392**, L9-L13 (1992).

22      Nicholl, M. *et al.* Slowly fading super-luminous supernovae that are not pair-instability explosions. *Nature* **502**, 346-349 (2013).

23      Greiner, J. *et al.* A very luminous magnetar-powered supernova associated with an ultra-long [ggr]-ray burst. *Nature* **523**, 189-192, doi:10.1038/nature14579 (2015).

24      Wheeler, J. C., Meier, D. L. & Wilson, J. R. Asymmetric Supernovae from Magnetocentrifugal Jets. *The Astrophysical Journal* **568**, 807-819 (2002).

25      Pessah, M. E. & Goodman, J. On the Saturation of the Magnetorotational Instability Via Parasitic Modes. *The Astrophysical Journal Letters* **698**, L72-L76 (2009).

26      Goodman, J. & Xu, G. Parasitic instabilities in magnetized, differentially rotating disks. *The Astrophysical Journal* **432**, 213-223 (1994).

27     Ott, C. D., Burrows, A., Thompson, T. A., Livne, E. & Walder, R. The Spin Periods and Rotational Profiles of Neutron Stars at Birth. *The Astrophysical Journal Supplement Series* **164**, 130-155 (2006).

**Acknowledgements** The authors would like to thank S. Couch, J. Zrake, D. Tsang, C. Wheeler, E. Bentivegna, and I. Hinder for discussions. This research was partially supported by NSF grants AST-1212170, PHY-1151197, OCI-0905046, and the Sherman Fairchild Foundation. The simulations were carried out on NSF/NCSA BlueWaters (PRAC ACI-1440083).

**Contributions P.M.:** Project planning, simulation code development, simulations, simulation analysis, visualisation, interpretation, manuscript preparation. **C.D.O.:** Group and project leadership, idea for the project, project planning, interpretation, manuscript preparation. **D.R.:** Simulation analysis, interpretation, simulation code development, manuscript preparation. **L.F.R.:** Interpretation, manuscript review. **E.S.:** Simulation code development, manuscript review. **R.H.:** Simulation code development, visualisation software development, manuscript review.

**Author Information**. Reprints and permissions information is available at www.nature.com/reprints. The authors declare that they have no competing financial interests. Correspondence and requests for materials should be addressed to Philipp Mösta. (email: pmoesta@tapir.caltech.edu).

**Figure 1:** Evolution of the maximum toroidal magnetic field. Both panels show the maximum toroidal magnetic field as a function of time for the four resolutions 500 m, 200 m, 100 m, and 50 m. The left panel shows the global maximum field, the right panel the maximum field in a thin layer above and below the equatorial plane (-7.5 km ≤ z ≤ 7.5 km). The magenta line indicates exponential growth with an e-folding time $\tau$ = 0.5 ms.

**Figure 2:** Radial magnetic field strength. Visualisation of the radial component of the magnetic field in 2D rz-slices at azimuth = 45° for the four resolutions 500 m, 200 m, 100 m, and 50 m at $t - t_{map}$ = 7.6 ms. The colourmap ranges from positive $10^{15}$ G (yellow) to negative $10^{15}$ G (light blue).

**Figure 3:** Turbulent kinetic and electromagnetic energy spectra. The top two panels show the energy as a function of dimensionless wavenumber k. The top left panel compares the electromagnetic energy across all four resolutions at $t - t_{map}$ = 10 ms (final simulated time). The top right panel shows a time series of electromagnetic energy spectra for the 50 m simulation only. The two upper panels show the turbulent kinetic energy as computed from the 50 m simulation, a line indicating Kolmogorov scaling ($k^{-5/3}$), and the initial electromagnetic energy spectrum. The bottom panel shows the electromagnetic energy at a given wavenumber $E_k$ versus time and an exponential and linear fit.

**Figure 4:** 3D volume renderings of the toroidal magnetic field. All panels show ray-casting volume renderings of $B^{\varphi}$. The rotation axis z is the vertical and the volume renderings are generated using a varying-alpha colour map. Yellow indicates positive field of strength $10^{15}$ G and red indicates weaker positive field. Light blue corresponds to negative field of $10^{15}$ G, while blue indicates weaker negative field. The left most panel shows the initial conditions for our simulations, the middle panel the 500 m simulation at time $t - t_{map} = 10$ ms and the right panel the 50m simulation at $t - t_{map} = 10$ ms.

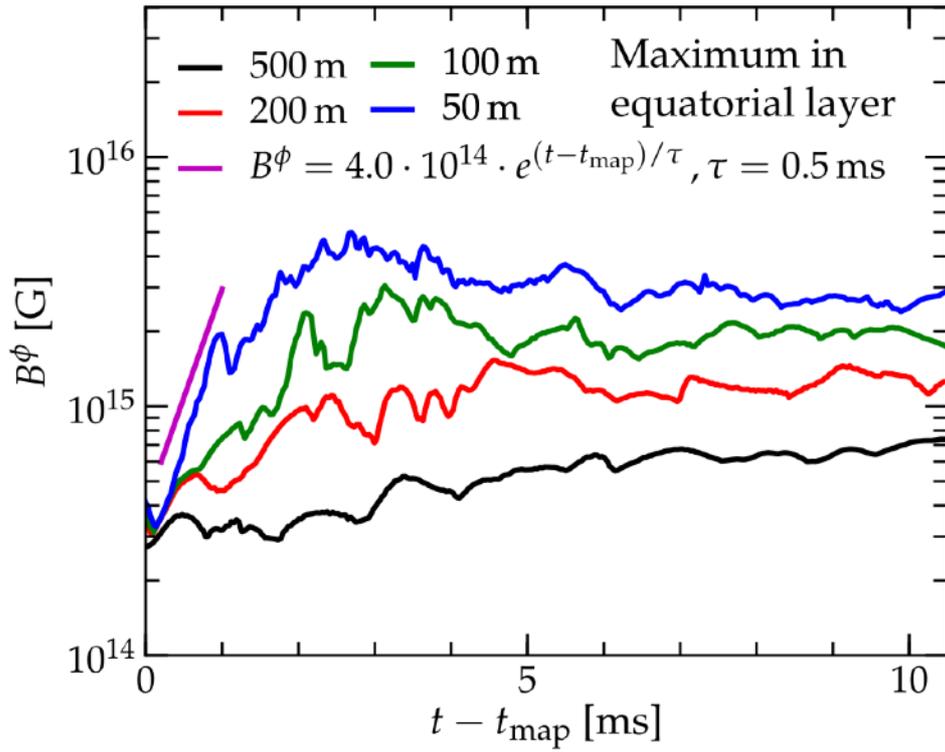

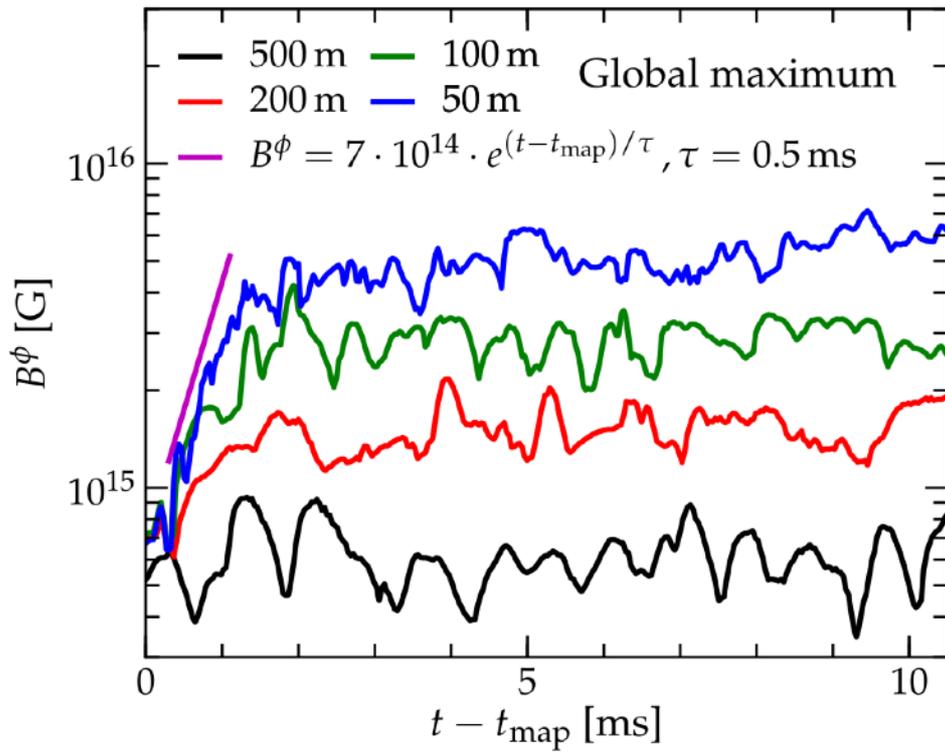

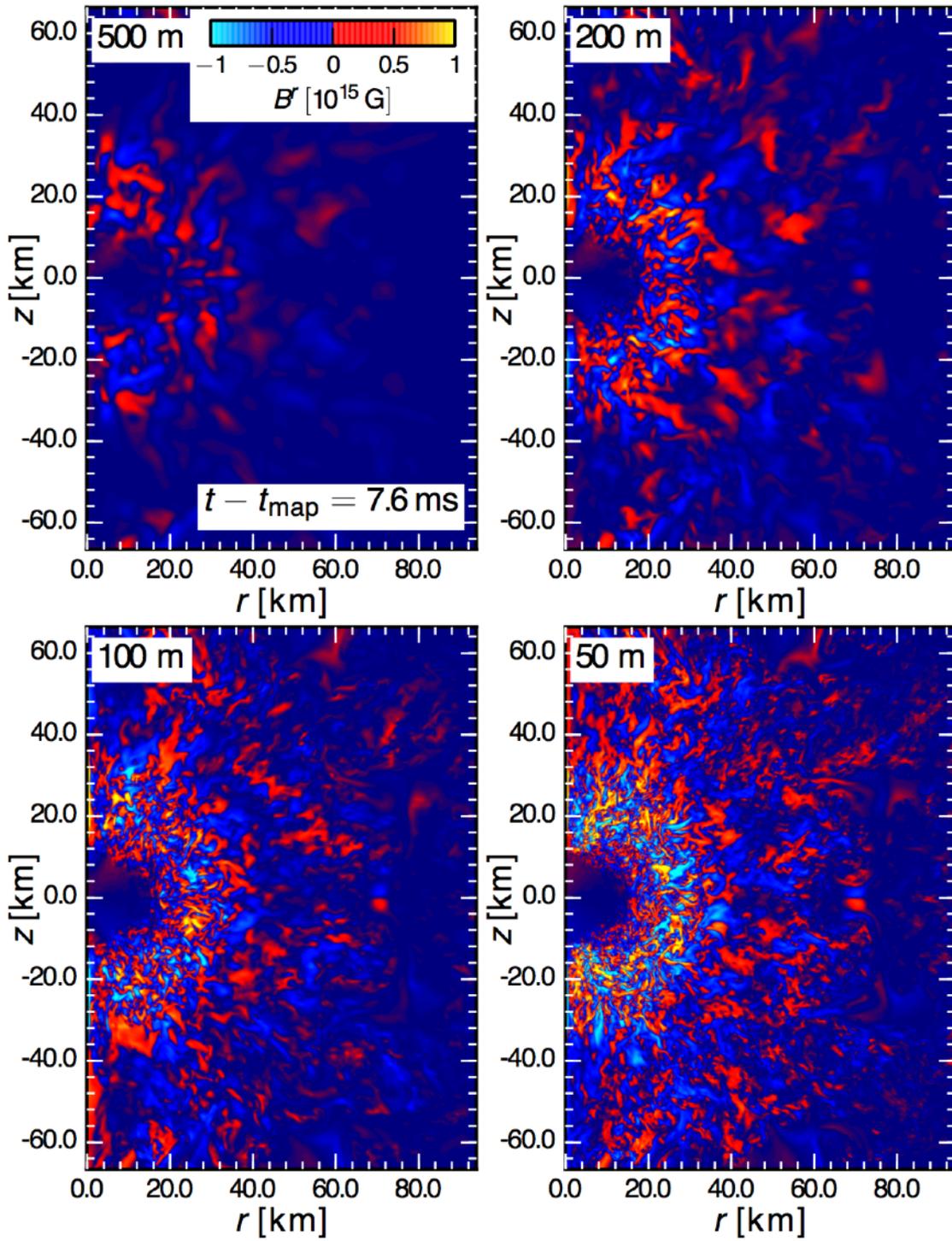

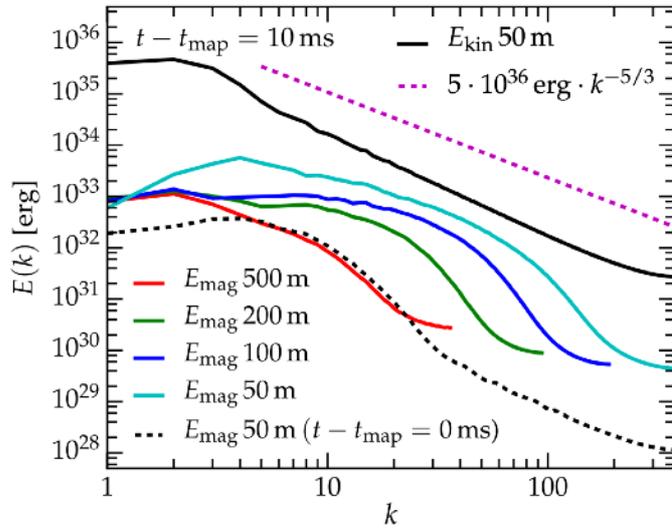

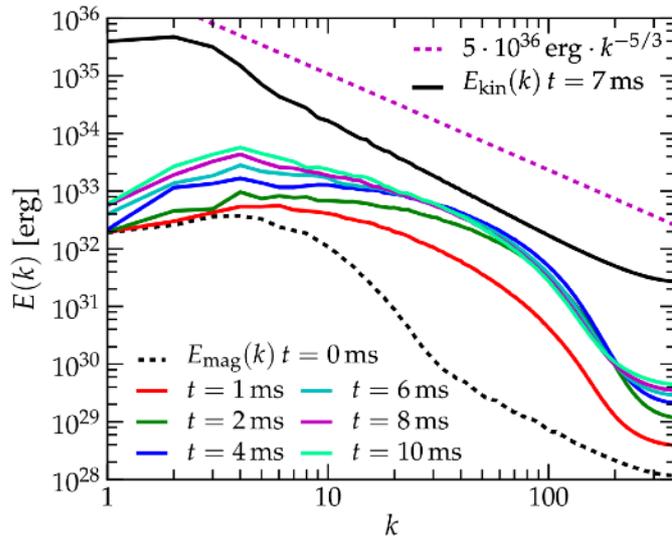

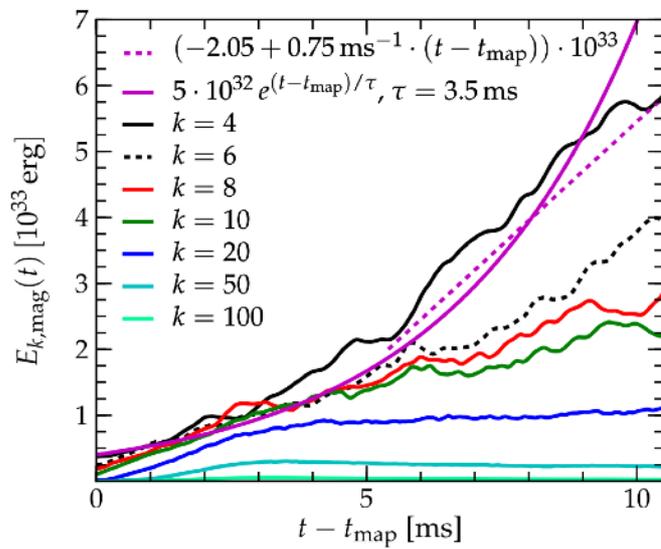

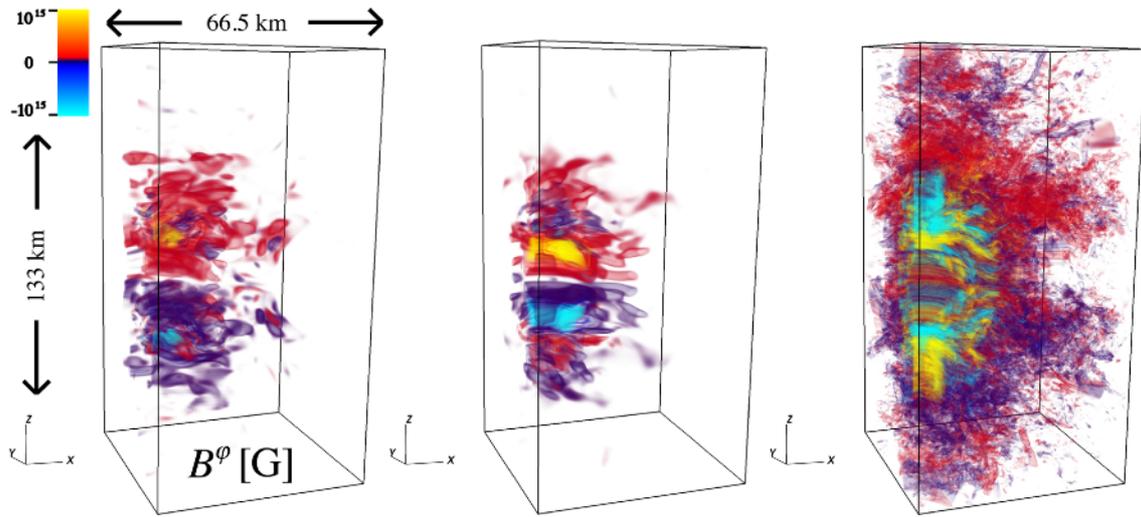

## Methods

## Initial conditions: Stellar collapse simulation

We start by performing a dynamical spacetime ideal MHD simulation with adaptive mesh refinement (AMR) of the 25-$M_\odot$ (at zero-age-main-sequence) presupernova model E25 from [28] with initial conditions for differential rotation as in [29] (initial central angular velocity of the iron core 2.8 rad $s^{-1}$, $x_0 = 500$ km and $z_0 = 2000$ km). This model could be considered as a type Ic-bl/hypernova and long gamma-ray burst progenitor[30]. At the onset of collapse, we set up a modified dipolar magnetic field structure from a vector potential of the form $A_r = A_\theta = 0$; $A_\varphi = B_0 (r_0^3) (r^3 + r_0^3)^{-1} r \sin \theta$, with $r_0 = 1000$ km as in [29], but with $B_0 = 10^{10}$ G. This progenitor seed field is not unreasonable for GRB supernova progenitor cores[30,31]. With the grid setup (9 levels of box-in-box AMR, finest resolution dx = 375 m) and methods identical to [29,32], we follow this simulation until $t_{map} = 20$ ms after core bounce. At this time, the initial supernova shockwave has stalled at a radius of $\simeq 130$ km. Radial profiles of key state variables (density, entropy, angular velocity, and fast magnetosonic speed) of the simulation at the time of mapping are shown in Extended Data Fig. 1 and 2. Both the protoneutron star and the post-shock region have reached a quasi-equilibrium state and the underlying spacetime changes only very slowly and secularly, which allows us to carry out subsequent high-resolution GRMHD simulations assuming a fixed background spacetime for $\sim 10 - 20$ ms. The resolution of the AMR box covering the shear layer of the protoneutron star in this initial simulation is dx = 750 m, but to resolve the fastest growing mode (FGM) of the MRI for the chosen initial magnetic field of $10^{10}$ G, a linear resolution of at least dx ~ 100 m is required[33]. This is why the common approach in the community to obtain the field strength to power a magnetorotational explosion ($\geq 10^{15}$ G) has been by flux-compression (B $\propto \rho^{2/3}$, amplification by a factor ~ $10^3$) from unrealistically high seed fields ($|B| \geq 10^{12}$ G precollapse)[34-37].

**Background flow stability analysis**

A magnetised fluid is unstable to weak-field shearing modes in the presence of a negative angular velocity gradient that is not compensated for by compositional or entropy gradients of the fluid[38]. At the time of mapping the initial AMR simulation to the high-resolution domain, the plasma in the shocked region around the protoneutron star is locally unstable to weak-field shearing modes where $C_{MRI} \equiv (\omega_{BV}^2 + r\, d\Omega^2/dr)/\Omega^2 < 0$[33,38,39]. Here $\omega_{BV}$ is the Brunt-Väisälä frequency indicating convective stability/instability, $r\, d\Omega^2/dr$ characterises the rotational shear, and $\Omega$ is the angular velocity. We follow [39,40] and calculate the stability criterion $C_{MRI}$, and the wavelength $\lambda_{FGM}$ and growth rate $\tau_{FGM}$ of the FGM of the MRI in 2D xy- and xz-slices through our 3D domain. To better approximate the background flow in our 3D AMR stellar collapse simulation, we average in space and time. We first carry out a spatial averaging step and calculate averaged versions of the state variables of our simulation at every timestep, e.g. the spatially averaged density $\rho_i$. For that, we choose a centered stencil taking into account 3 points in each direction (this is the maximum number of points we have available at AMR component boundaries). Since this is insufficient to get a large enough sample of points for the averaging procedure, we additionally calculate a moving time average of the form $\rho_{av,i} = \alpha \cdot \rho_i + (1 - \alpha) \cdot \rho_{av,i-1}$, where i denotes the current timestep and i − 1 the previous one. We choose a weight function for each dataset in the moving average as $\alpha = 2 \cdot (\Delta t/\Delta t_{coarse} \cdot n + 1)^{-1}$, where $\Delta t$ is the timestep on the current refinement level and $\Delta t_{coarse}$ the timestep of the coarsest level. This choice of weight function guarantees that 86% of the data in the average is comprised of the last n timestep datasets. The timestep size in our AMR simulation on the refinement level containing the shear layer around the protoneutron star is $\Delta t = 5 \times 10^{-4}$ ms and we choose n such that $\alpha = 2000$, ensuring temporal averaging over a timescale of $\simeq 1$ ms. We calculate $C_{MRI}$, $\lambda_{FGM}$, and $\tau_{FGM}$ from the space and time averages of the state variables in our simulation (Extended Data Fig. 3).

**Mapping to high-resolution computational domain**

Next, we map the configuration to a 3D domain with uniform spacing of the form x, y, z = [−66.5 km, 66.5 km] for four resolutions h = [500 m, 200 m, 100 m, 50 m]. To guarantee divergence-free initial data for the magnetic field, we carry out a constraint projection step after we have interpolated the magnetic field to the new domain. This is technically challenging as we have to make sure that all operators used in the projection are consistent in their definition with the discrete form of the divergence operator maintained in our specific implementation of constrained transport[32]. We use a discrete analog of the Helmholtz decomposition[40] to decompose the magnetic field into a discrete curl $\text{curl}_h$ and a discrete gradient $\text{grad}_h$,

$$B = \text{curl}_h A + \text{grad}_h \Phi , \tag{1}$$

where $\Phi$ is a discrete scalar field. The discrete divergence $\text{div}_h$ of (1) leads to a discrete Poisson equation

$$\text{div}_h B = \Delta_h \Phi , \tag{2}$$

where $\Delta_h$ is the discrete Laplace operator. We solve (2) augmented with homogeneous Dirichlet boundary conditions to machine precision for $\Phi$ using the conjugate gradient solver provided by the PETSc[41] library in combination with the parallel algebraic multi-grid preconditioner HYPRE[42]. We then obtain a divergence free field $B'$ from the projection

$$B' = B - \text{grad}_h \Phi. \tag{3}$$

Finally, we recompute $\text{div}_h B'$ to check that it is zero to floating point precision.

**High-resolution turbulence simulations**

We perform ideal, fixed background spacetime, GRMHD simulations using the open-source Einstein Toolkit[32,41] with WENO5 reconstruction[42,43], the HLLE Riemann solver[44] and constrained transport[45] for maintaining $\text{div}B = 0$. We employ the $K_0 = 220$ MeV variant of the finite-temperature nuclear equation of state of [46] and the neutrino leakage/ heating approximations described in [47] and [48] with a heating scale factor $f_{heat} = 1.0$. We perform simulations on a domain with uniform spacing of the form x, y = [0 km, 66.5 km] and z = [−66.5 km, 66.5 km] for four resolutions h = [500 m, 200 m, 100 m, 50 m] in quadrant symmetry 3D (90-degree rotational symmetry in the xy-plane). We keep all variables at the boundary fixed in time. This is justified for a number of reasons. First, the accretion boundary flow itself only changes on timescales longer than those simulated. Secondly, the fast magnetosonic speed (Extended Data Fig. 1 panel (d) and Extended Data Fig. 2, panel (c)) is of the order of a few percent of the speed of light throughout the high-resolution computational domain. This implies a boundary crossing time for the simulation box of ∼ 20 ms. This leaves the results in the shear layer unaltered by boundary effects for the simulated times of 10 ms. Additionally, as the cylindrically rotating flow in the shocked region is rotating in and out of the purely Cartesian boundary zones sounds waves can be reflected at the boundaries. While these reflections are not necessarily unrealistic, as there will be perturbations in the shocked region of any rotating iron core, they pose an additional complication for the numerical stability of the simulations[49]. We find these reflections to be minimal in the hydrodynamical variables themselves, but they do cause spurious oscillations in the magnetic field towards the boundary zones. To prevent these oscillations at the outer boundary without affecting the solution in the shear layer around the protoneutron star, we apply diffusivity at the level of the induction equation for the magnetic field via a modified Ohm's law. We choose $\mathbf{E} = -\mathbf{v} \times \mathbf{B} + \eta \mathbf{J}$, where $\mathbf{J} = \nabla \times \mathbf{B}$ is the 3-current density and set $\eta = \eta_0 \cdot (0.5 + 0.5 \tanh ((r

$- r_{\text{diff}}$ ) $b^{-1}$)) with $\eta_0 = 10^{-2}$, $r_{\text{diff}} = 40$ km and $b = 3$ km. That is, we apply diffusivity only in a region outside of radius $r_{\text{diff}}$ and transition smoothly over a blending zone with width b to no diffusivity inside $r_{\text{diff}}$.

**Turbulent kinetic and magnetic energy spectra**

We compute spectra of the turbulent kinetic and magnetic energy as instantaneous snapshots using the discrete Fourier transform $\hat{u}(k) = \sum_x u(x) \exp(-2\pi i \frac{kx}{L})(\frac{L}{N})^3$ [50]. The spectra shown in Fig. 3 are densitised to better reflect the overall energy contained in the turbulent kinetic motion and the magnetic field. In Extended Data Fig. 5, we show the spectra of the non-densitised turbulent velocity (panel (a)) and the non-densitised magnetic field (panel (b)) and additionally window the data to account for the non-periodicity at the boundaries of our computational domain. For that we use a mollifier of the form $m(x) = \exp(-1/(1 - (\frac{abs((x-d))}{d})^2) + 1)$, and respectively for y, and z.

This effectively blends the data to zero over a stencil width d at the outer boundary. We choose $d = 3$, but note that other choices yield similar results. These non-densitised and windowed spectra illustrate that the lack of an exponential turnoff at large k in the turbulent kinetic energy in Fig. 3 is due to the inclusion of the nearly discontinuous density falloff at the edge of the protoneutron star core (at $r \simeq 12$ km) in the calculation of the spectrum for Fig. 3 and the non-periodicity of our computational domain. The non-densitised and windowed turbulent kinetic energy spectrum in Extended Data Fig. 5 is compensated for $k^{-5/3}$ scaling (as expected in Kolmogorov theory). We observe a slightly steeper scaling between $k^{-5/3}$ and $k^{-2}$. Within the first 3 ms there is a rapid transition into a fully turbulent state at large k (panel (b), Fig. 3, panel (a) Extended Data Fig. 5). Afterwards, the turbulent kinetic energy decreases at large k and the spectrum gradually evolves towards a steeper falloff. There is no increase in the turbulent kinetic energy at small k at late times. The magnetic energy, similarly to the turbulent kinetic energy, peaks at large k at t - $t_{map}$ ≈ 3 ms, which correlates well with the observed saturation of the maximum toroidal field shown in Fig. 1. Subsequently, the magnetic energy at small k grows first exponentially and then linearly with time. This picture is consistent with energy being extracted from the turbulent kinetic motion at large k and being pumped into an inverse cascade that leads to magnetic field energy growth at small k. As the kinematic phase ends and transitions into saturation, magnetic fields and numerical resistivity become important for the evolution[51]. This may explain the transition to linear growth. We additionally observe a superposed 2 ms modulation on top of the k = 4 exponential growth that corresponds roughly to the Alfven crossing time across the shear layer ($t_{A,shear}$~2 ms).

**Angle-averaged magnetic flux and poloidal current**

We compute the 2D angle-averaged (in $\varphi$) magnetic flux and poloidal current to determine which magnetic field structures are global in $\varphi$ (Extended Data Fig. 6 and

Extended Data Fig. 7). The magnetic flux is computed as $\int_0^{\omega_{max}} \omega\, B^z\, d\omega$ and the current as $J = \nabla \times B$. The iso-contours of the magnetic flux represent the poloidal field lines, while the poloidal current approximates the toroidal magnetic field. We find that the shear layer of the protoneutron star distorts the initial poloidal magnetic field of the iron core, but we find no emerging global poloidal field created from turbulence. The toroidal field (poloidal current) however does show a global structure that roughly fills the width of the shear layer in the polar region of our simulation supporting the idea that toroidal magnetar-strength field in our simulations (see also Fig. 4) truly is global in $\varphi$.

**Limitations of this study**

Limitations of this study are finite resolution (most visible in the not fully converged saturation magnetic field) and the sensitivity of the detailed turbulent state on the numerical methods. Additionally, the impact of the imposed 90–degree rotational symmetry will have to be investigated. Ultimately, the high-resolution simulations of the kind performed here have to be embedded back into a full-star simulation to determine the detailed shock revival and explosion geometry.

**Code availability** All computer code used in this study that is not already open source will be made available under http://stellarcollapse.org

**References**

28    Heger, A. & Langer, N. Presupernova Evolution of Rotating Massive Stars. II. Evolution of the Surface Properties. *The Astrophysical Journal* **544**, 1016-1035 (2000).

29    Mösta, P. *et al.* Magnetorotational Core-collapse Supernovae in Three Dimensions. *The Astrophysical Journal Letters* **785**, L29 (2014).

30    Woosley, S. E. & Heger, A. The Progenitor Stars of Gamma-Ray Bursts. *The Astrophysical Journal* **637**, 914-921 (2006).

31    Wheeler, J. C., Kagan, D. & Chatzopoulos, E. The Role of the Magnetorotational Instability in Massive Stars. *The Astrophysical Journal* **799**, 85 (2015).

32    Mösta, P. *et al.* GRHydro: a new open-source general-relativistic magnetohydrodynamics code for the Einstein toolkit. *Classical and Quantum Gravity* **31**, 5005 (2014).

33    Obergaulinger, M., Aloy, M. A. & Müller, E. Axisymmetric simulations of magneto-rotational core collapse: dynamics and gravitational wave signal. *Astronomy and Astrophysics* **450**, 1107-1134 (2006).

34    Burrows, A., Dessart, L., Livne, E., Ott, C. D. & Murphy, J. Simulations of Magnetically Driven Supernova and Hypernova Explosions in the Context of Rapid Rotation. *The Astrophysical Journal* **664**, 416-434 (2007).

35    Takiwaki, T. & Kotake, K. Gravitational Wave Signatures of Magnetohydrodynamically Driven Core-collapse Supernova Explosions. *The Astrophysical Journal* **743**, 30 (2011).

36    Winteler, C. *et al.* Magnetorotationally Driven Supernovae as the Origin of Early Galaxy r-process Elements? *The Astrophysical Journal Letters* **750**, L22 (2012).

37    Nishimura, N., Takiwaki, T. & Thielemann, F.-K. The r-process nucleosynthesis in the various jet-like explosions of magnetically driven core-collapse supernovae. *ArXiv e-prints* **1501**, 6567 (2015). <http://adsabs.harvard.edu/abs/2015arXiv150106567N%3E.

38    Balbus, S. A. & Hawley, J. F. A powerful local shear instability in weakly magnetized disks. I - Linear analysis. II - Nonlinear evolution. *The Astrophysical Journal* **376**, 214-233 (1991).

39    Balbus, S. A. & Hawley, J. F. Instability, turbulence, and enhanced transport in accretion disks. *Reviews of Modern Physics* **70**, 1-53 (1998).

40    Akiyama, S., Wheeler, J. C., Meier, D. L. & Lichtenstadt, I. The Magnetorotational Instability in Core-Collapse Supernova Explosions. *The Astrophysical Journal* **584**, 954-970 (2003).

41    Löffler, F. *et al.* The Einstein Toolkit: a community computational infrastructure for relativistic astrophysics. *Classical and Quantum Gravity* **29**, 5001 (2012).

42    Reisswig, C. *et al.* Three-dimensional general-relativistic hydrodynamic simulations of binary neutron star coalescence and stellar collapse with multipatch grids. *Physical Review D* **87**, 64023 (2013).

43    Tchekhovskoy, A., McKinney, J. C. & Narayan, R. WHAM: a WENO-based general relativistic numerical scheme - I. Hydrodynamics. *Monthly Notices of the Royal Astronomical Society* **379**, 469-497 (2007).

44      Einfeldt, B. in *Shock Tubes and Waves* (ed Hans Groenig) 671-676 (1988).

45      Tóth, G. The $\nabla \cdot B=0$ Constraint in Shock-Capturing Magnetohydrodynamics Codes. *Journal of Computational Physics* **161**, 605-652 (2000).

46      Lattimer, J. M. & Douglas Swesty, F. A generalized equation of state for hot, dense matter. *Nuclear Physics A* **535**, 331-376 (1991).

47      O'Connor, E. & Ott, C. D. A new open-source code for spherically symmetric stellar collapse to neutron stars and black holes. *Classical and Quantum Gravity* **27**, 4103 (2010).

48      Ott, C. D. *et al.* Correlated gravitational wave and neutrino signals from general-relativistic rapidly rotating iron core collapse. *Physical Review D* **86**, 24026 (2012).

49      Bogovalov, S. V. Boundary conditions and critical surfaces in astrophysical MHD winds. *Astronomy and Astrophysics* **323**, 634-643 (1997).

50      Eswaran, V. & Pope, S. B. An examination of forcing in direct numerical simulations of turbulence. *Computers & Fluids* **16**, 257-278 (1988).

51      Zrake, J. & MacFadyen, A. I. Magnetic Energy Production by Turbulence in Binary Neutron Star Mergers. *The Astrophysical Journal Letters* **769**, L29 (2013).

**Extended Data Figure 1:** AMR stellar collapse simulation. All panel show profiles along the x-direction of the initial stellar collapse simulation 20 ms after core bounce. Panel a) shows density, panel b) entropy, panel c) the angular velocity, and panel d) the fast magnetosonic speed.

**Extended Data Figure 2:** AMR stellar collapse simulation. All panel show profiles along the z-direction of the initial stellar collapse simulation 20 ms after core bounce. Panel a) shows density, panel b) entropy, and panel c) the fast magnetosonic speed.

**Extended Data Figure 3:** Background flow stability analysis. The top two panels show the stability criterion $C_{MRI}$ 20 ms after core bounce for the initial stellar collapse simulation. The top left panel shows a 2D xy-slice (z=0) through the 3D domain, the top

right panel a xz-slice (y=0). Yellow and red indicate regions, which are stable to shearing modes, while dark and light blue colours indicate unstable regions. The bottom left panel shows the wavelength $\lambda_{FGM}$ of the FGM of the MRI, the bottom right panel the growth rate of the FGM $\tau_{FGM}$. Both lower panels are zoomed in on the shear layer around the protoneutron star.

**Extended Data Figure 4:** Evolution of the maximum poloidal magnetic field. Both panels show the maximum poloidal magnetic field as a function of time for the four resolutions 500 m, 200 m, 100 m, and 50 m. The left panel shows the global maximum field, the right panel the maximum field in a thin layer above and below the equatorial plane (-7.5 km ≤ z ≤ 7.5 km). The magenta line indicates exponential growth with an e-folding time $\tau_{FGM}$ = 0.5 ms.

**Extended Data Figure 5:** Non-densitised turbulent kinetic and electromagnetic energy spectra. Panel a) shows a time-series of non-densitised turbulent kinetic energy spectra as a function of dimensionless wavenumber k. A line indicating Kolmogorov scaling ($k^{-5/3}$), is shown. Panel b) shows a time-series of non-densitised magnetic energy spectra as a function of dimensionless wavenumber k. In both panels the initial spectrum at t - $t_{map}$ = 0 ms (dashed black line) is shown for reference.

**Extended Data Figure 6:** Angle-averaged poloidal magnetic current and magnetic flux. All panels show r-z slices (cylindrical coordinates, angle-averaged in $\varphi$) of the poloidal magnetic current (colour-coded) and superposed contours of magnetic flux (black lines)

at t - $t_{map}$ = 10.3 ms (final simulated time). Panel a) shows the 500 m simulation, panel b) the 200 m simulation, panel c) the 100 m simulation, and panel d) the 50 m simulation.

**Extended Data Figure 7:** Angle-averaged poloidal magnetic current and velocity vectors. All panels show r-z slices (cylindrical coordinates, angle-averaged in $\varphi$) of the poloidal magnetic current (colour-coded) and superposed velocity vectors (red arrows) at t - $t_{map}$ = 10.3 ms (final simulated time).

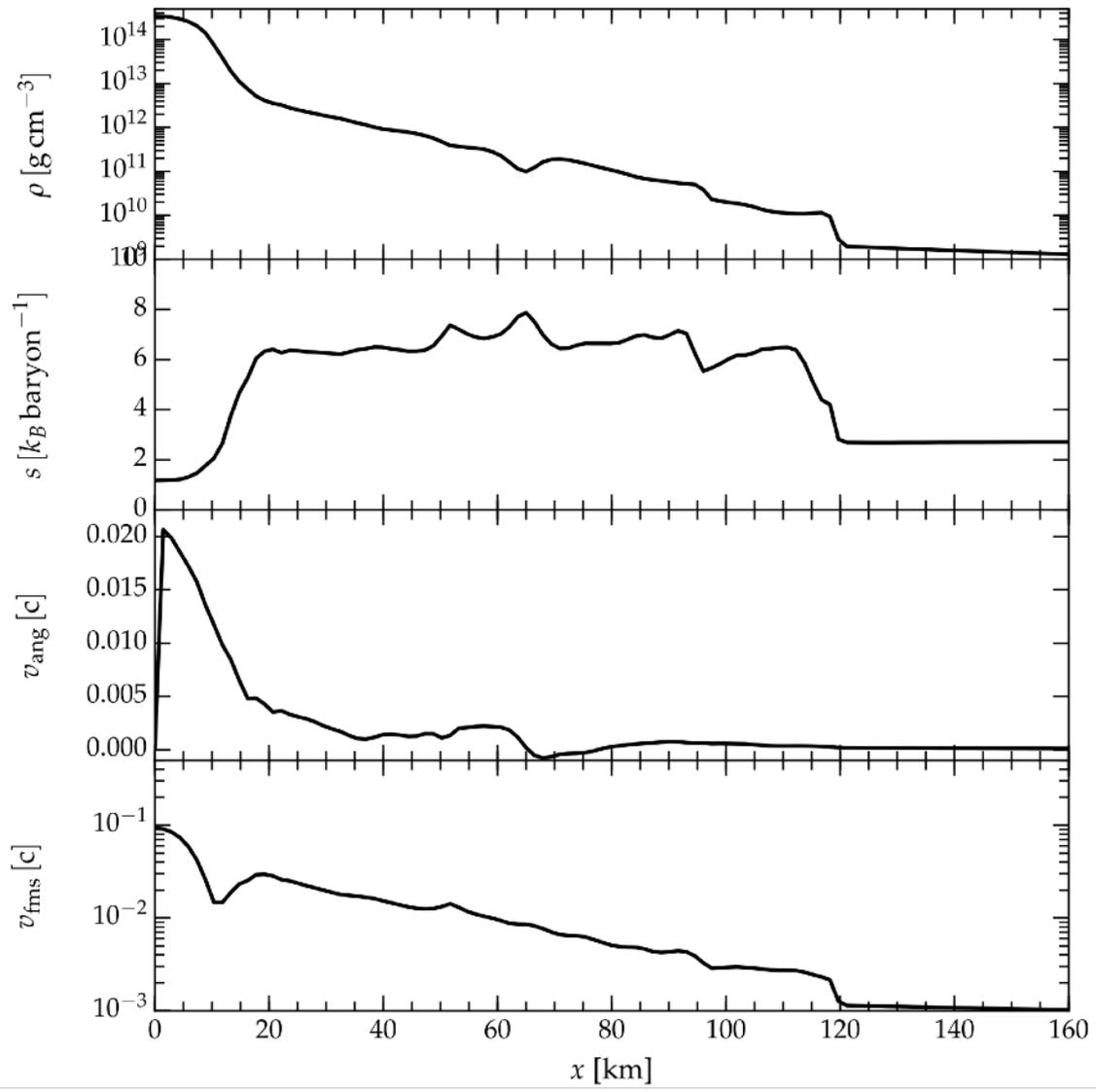

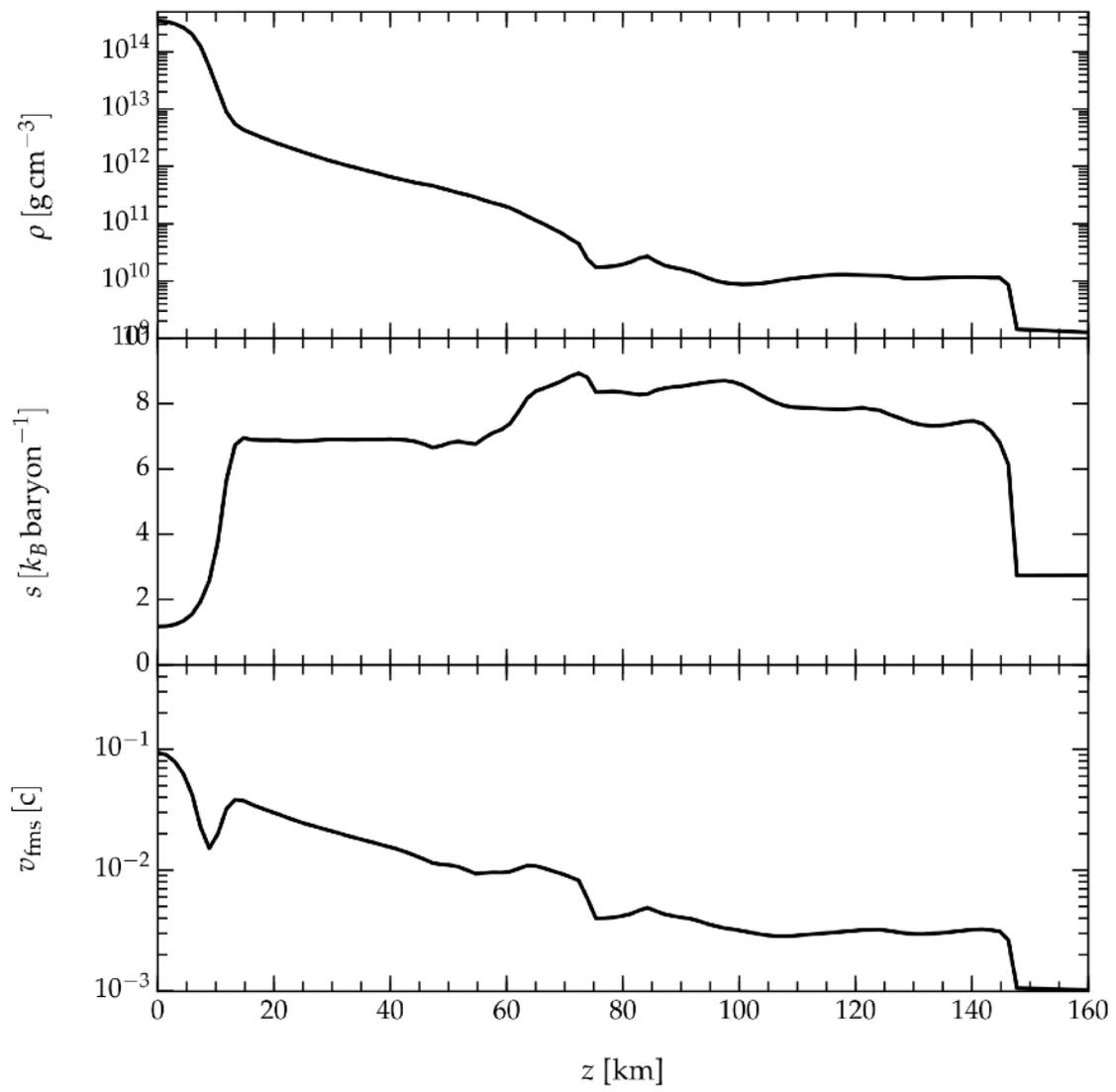

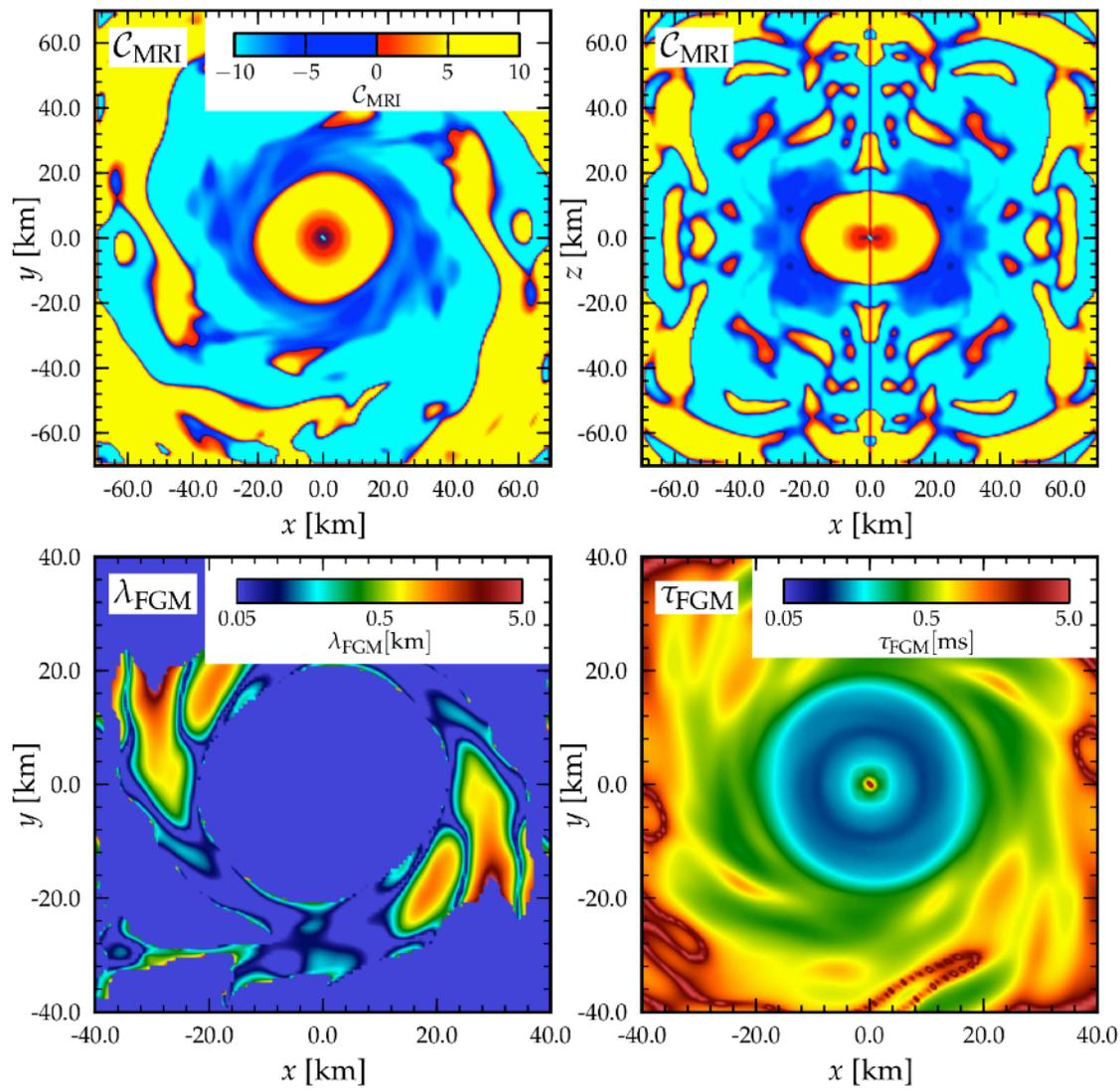

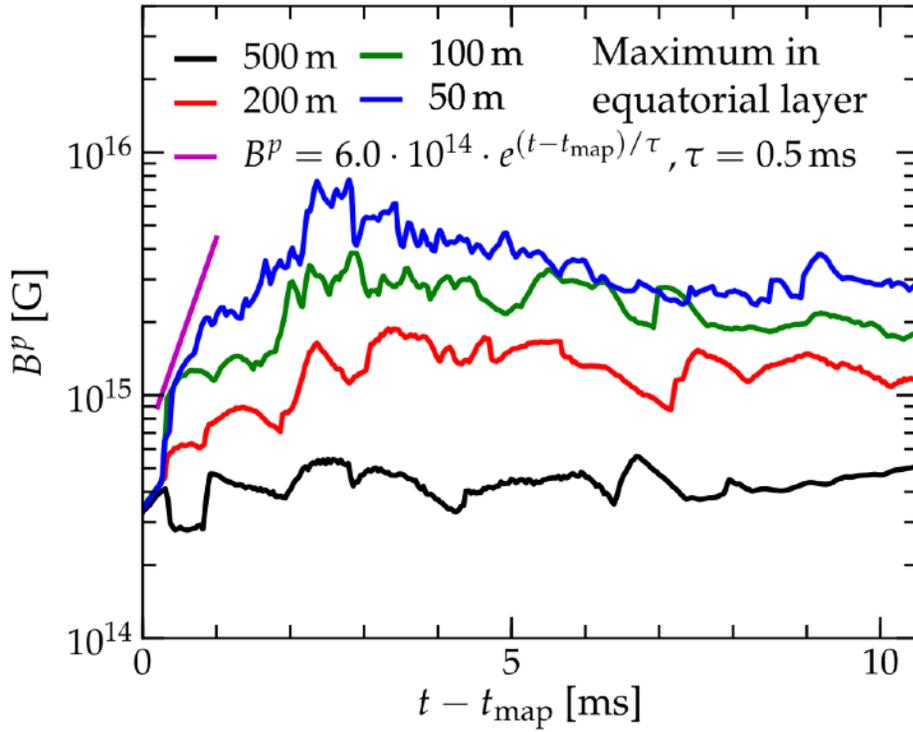

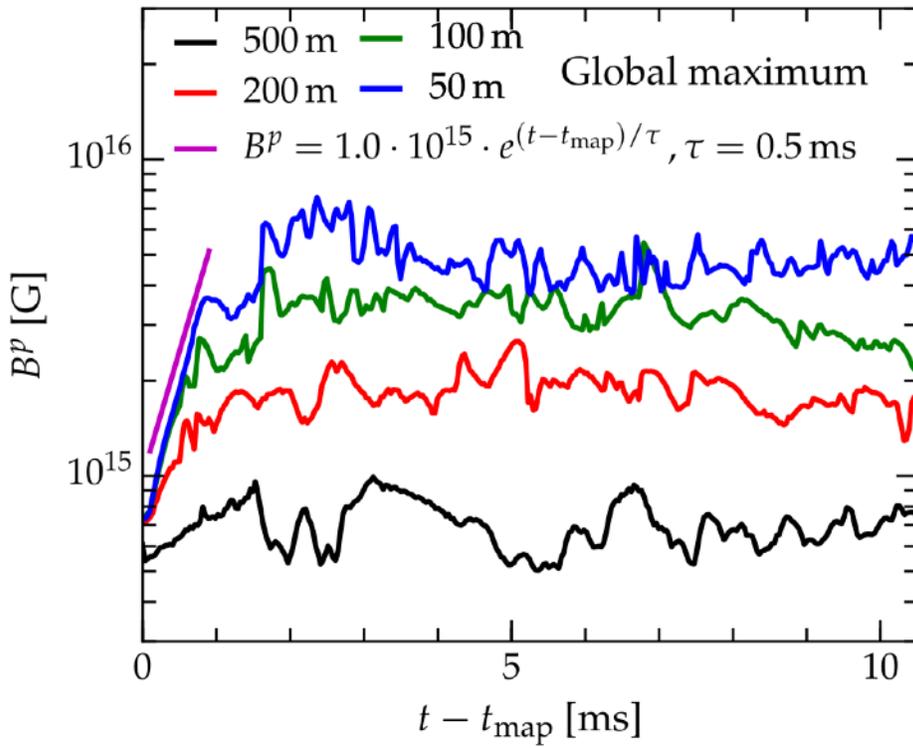

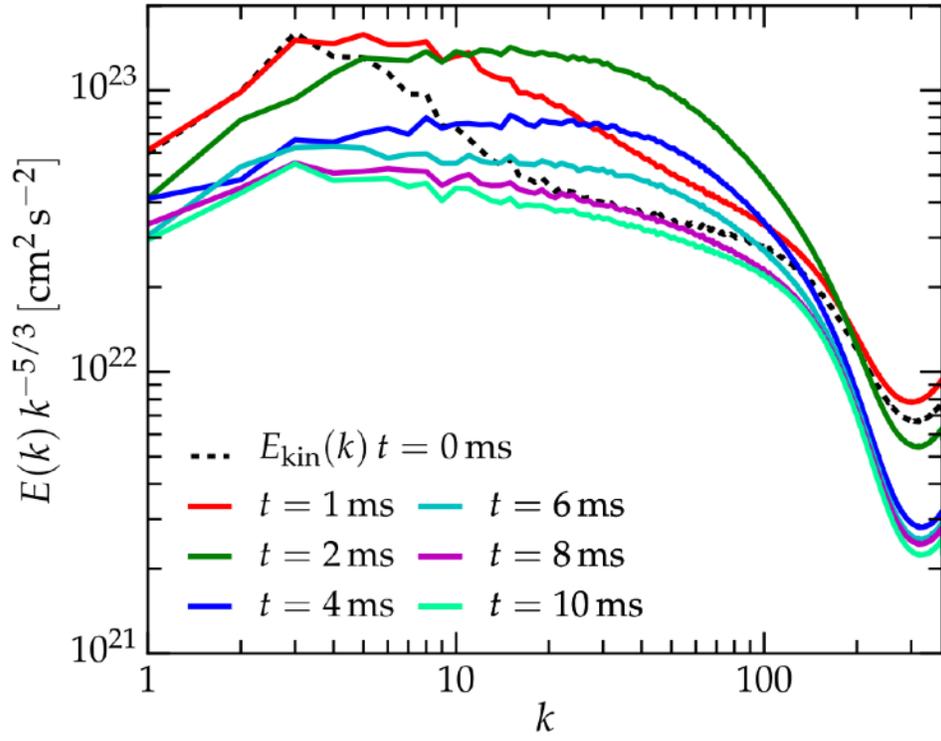

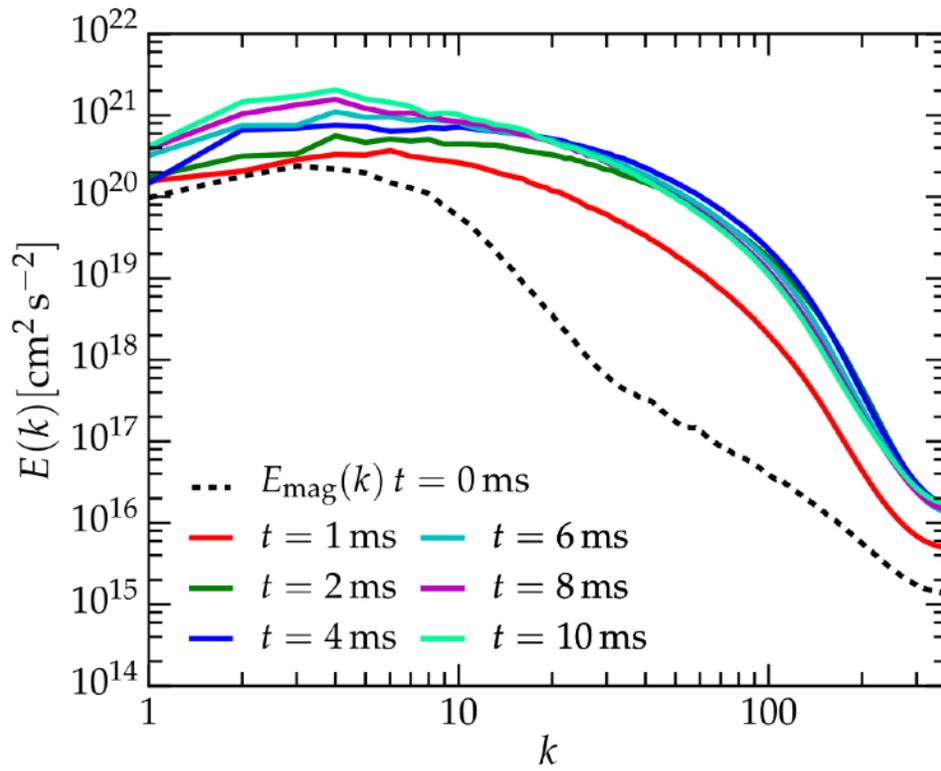

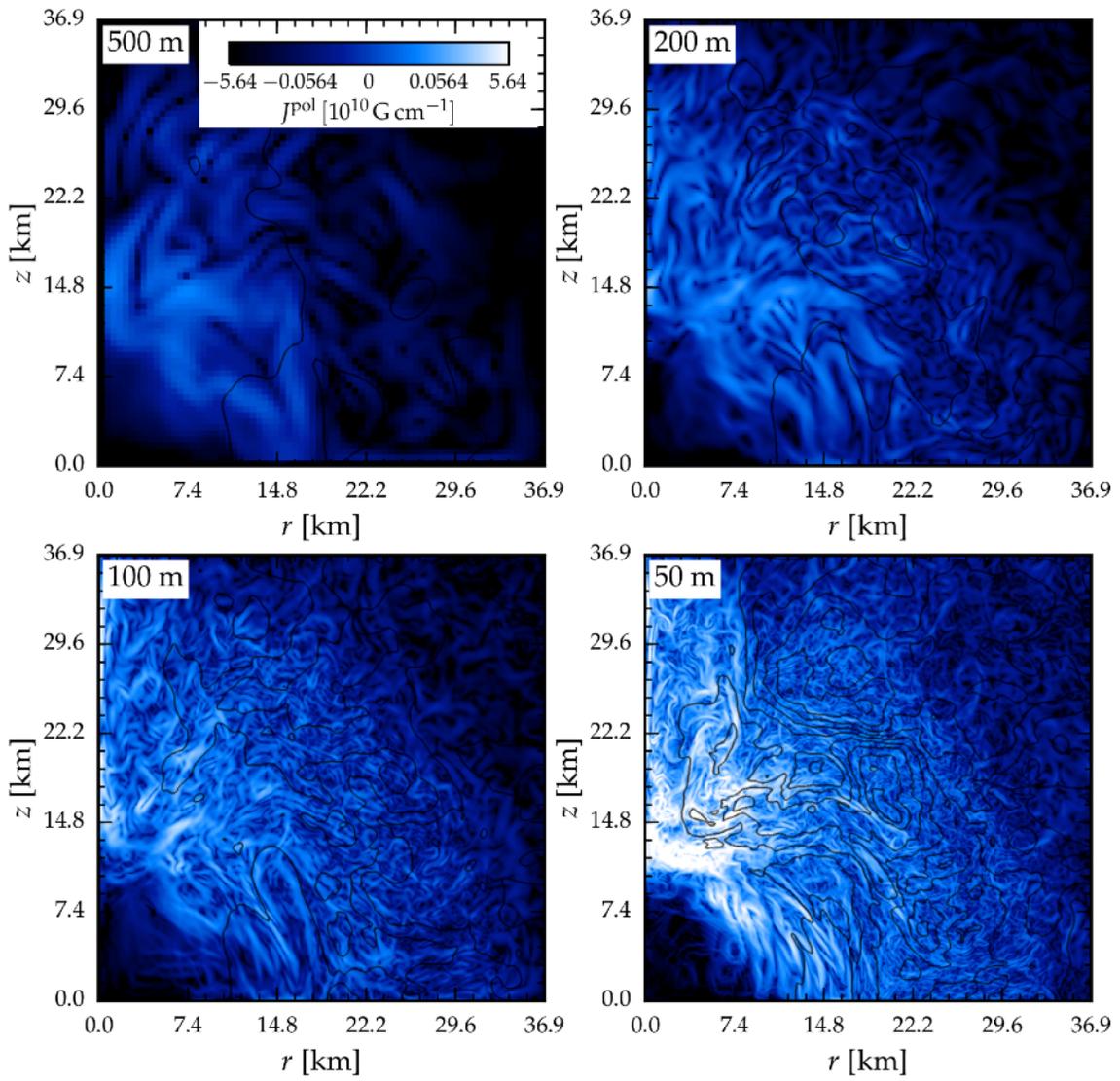

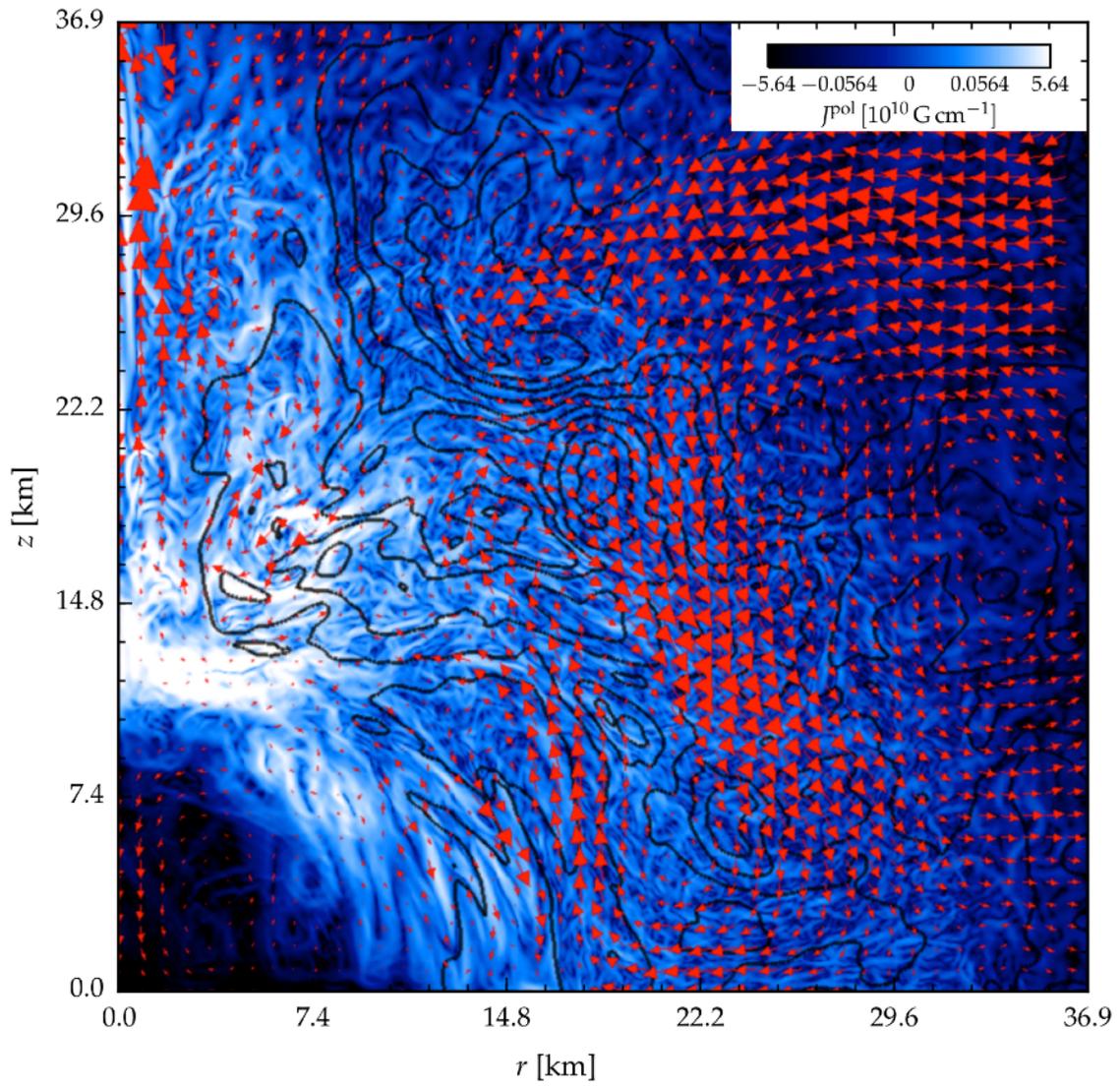